\documentclass[prx,twocolumn,superscriptaddress,nopacs,amsmath,amssymb]{revtex4-1}
\usepackage{graphicx}

\usepackage{hyperref}
\usepackage{color}
\usepackage{booktabs}
\usepackage{makecell}
\usepackage{ulem}

\usepackage[dvipsnames]{xcolor}

\begin{document}

\title{Topological Flat Bands and Correlated States in Twisted Monolayer-Bilayer Graphene}
\author{Louk Rademaker}
\affiliation{Department of Theoretical Physics, University of Geneva, 1211 Geneva, Switzerland}
\author{Ivan V. Protopopov}
\affiliation{Department of Theoretical Physics, University of Geneva, 1211 Geneva, Switzerland}
\author{Dmitry A. Abanin}
\affiliation{Department of Theoretical Physics, University of Geneva, 1211 Geneva, Switzerland}
\date{\today}

\begin{abstract}

Monolayer graphene placed with a twist on top of AB-stacked bilayer graphene hosts topological flat bands in a wide range of twist angles. The dispersion of these bands and gaps between them can be efficiently controlled by a perpendicular electric field, which induces topological transitions accompanied by changes of the Chern numbers. In the regime where the applied electric field induces gaps between the flat bands, we find a relatively uniform distribution of the Berry curvature. Consequently, interaction-induced valley- and/or spin-polarized states at integer filling factors are energetically favorable. In particular, we predict a quantum anomalous Hall state at filling factor $\nu=1$ for a range of twist angles $1^\circ<\theta <1.4^\circ$. Furthermore, to characterize the response of the system to magnetic field, we computed the Hofstadter butterfly and the Wannier plot, which can be used to probe the dispersion and topology of the flat bands in this material. 
\end{abstract}

\maketitle

Studies of twisted two-dimensional materials have gained an enormous boost after the discovery of correlated insulator states and superconductivity in twisted bilayer graphene (tBG)~\cite{Cao:2018ff,Cao:2018kn,Cao:2019ui,MacDonald:2019tm,Kerelsky:2019na,Yankowitz:2018tx}. tBG has been predicted~\cite{Bistritzer:2011ho} to be a natural platform for interaction physics due to the presence of magic angles where nearly flat bands near charge neutrality are formed. Following these results, a variety of related stacked materials with flat bands have been proposed and experimentally studied, including twisted double graphene bilayers~\cite{Cao:2019vx}, trilayer graphene on hBN~\cite{Chen:2019tf,Chen:2019vq,Chittari:2019fb}, and twisted transition-metal dichalcogenides~\cite{Wang:2019ux}. 

Several twisted materials naturally lead to topologically nontrivial band structures. Standard tBG is one of the first materials that display ``fragile topology''\cite{Po:2017bn,Po:2018vk,Song:2018ul,AhnFragile}. Trilayer graphene on hBN can exhibit nontrivial Chern bands under an applied field~\cite{Chen:2019tf,Chen:2019vq,Chittari:2019fb}. Further, recent experiments observed striking signatures of the quantum anomalous Hall (QAH) effect in tBG~\cite{Sharpe:vc,Serlin:2019ub}. Theoretical works~\cite{Bultinck:2019wp,2019arXiv191102045B,Zhang:2019hm} have argued that such a state may arise as a combined effect of alignment with the hBN substrate and Coulomb interaction, via a mechanism reminiscent of the quantum Hall ferromagnetism. 

\begin{figure}[t]
	\includegraphics[width=\columnwidth]{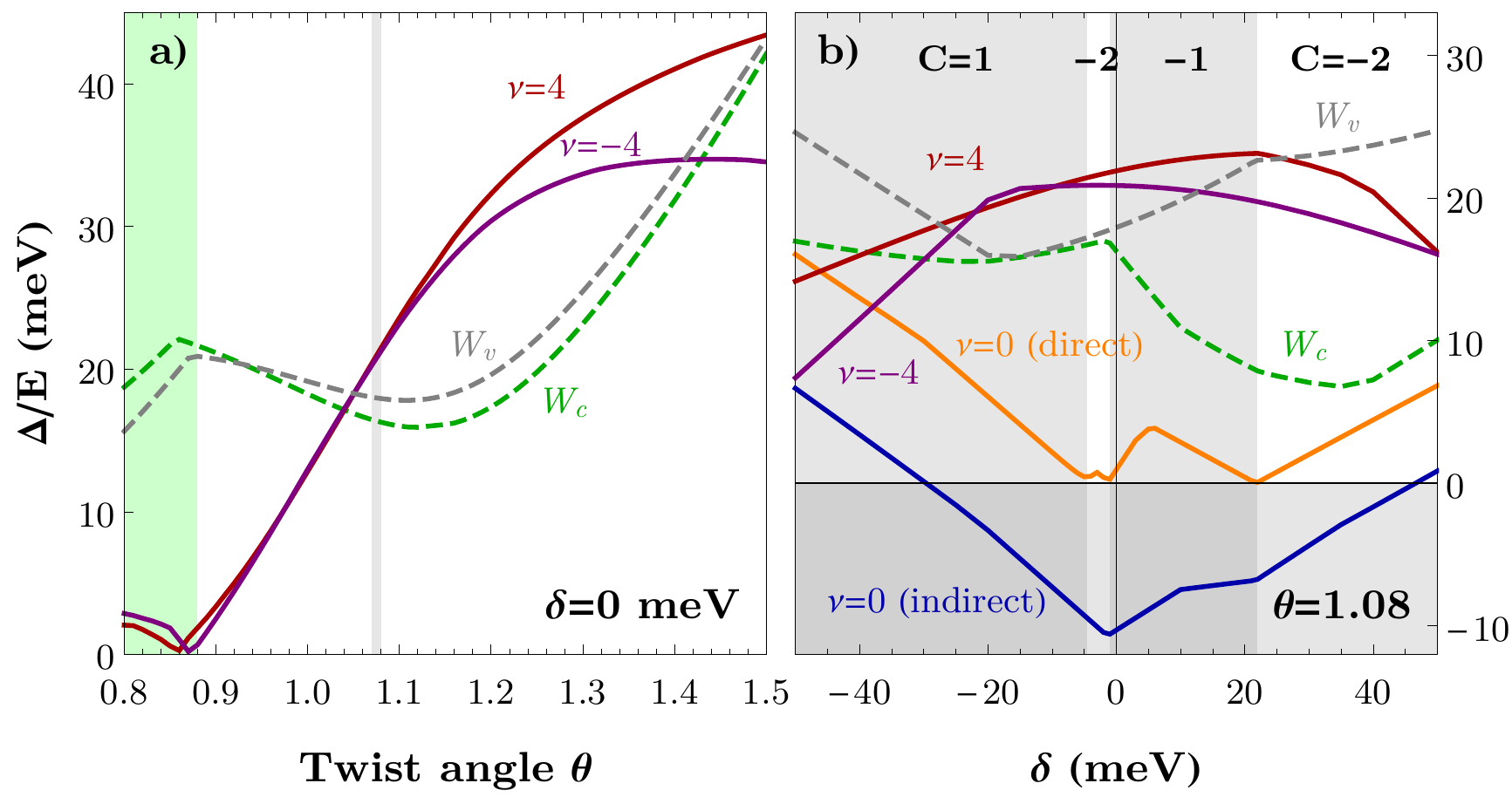} \\
	\includegraphics[width=\columnwidth]{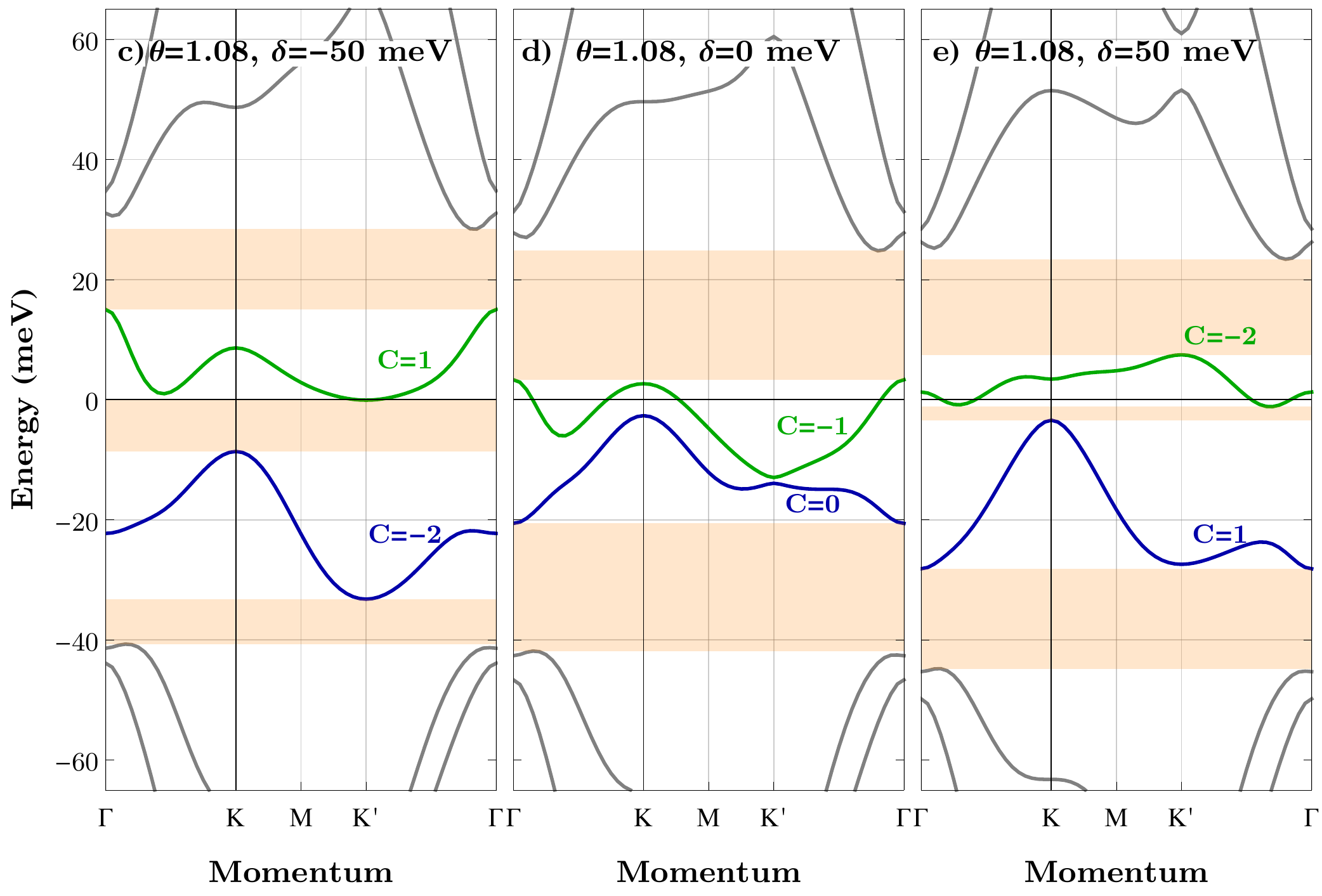}
	\caption{
	Twisted monolayer - bilayer graphene is a topologically nontrivial flat band system.
	{\bf a.} The bandwidth of the flat bands (dashed lines, $W_c$ for upper and $W_v$ for lower flat band) and the superlattice gaps (at fillings $\nu = -4$ and $\nu = 4$) as a function of twist angle. Below $\theta = 0.88^\circ$, the flat bands are no longer separated from the other bands. 
	{\bf b.} The flat bands can be tuned by applying a perpendicular electric field $\delta$, which causes a sequence of topological transitions. The Chern number of the upper flat band in the valley $\xi=1$ goes from $C=1$ at large negative fields to $C=-2$ at large positive fields. We also plot the direct and indirect gap between the two bands (at filling $\nu=0$), revealing that for large fields the two flat bands are well separated without closing the superlattice gaps at $\nu \pm 4$.
    {\bf c-e.} The dispersion of the flat bands in one valley ($\xi=1$) for twist angle $\theta = 1.08^\circ$ and various displacement field $\delta$. The gaps are indicated by a shaded region. Notice that the total valley Chern number remains constant, consistent with~\cite{LiuPRX2019}.}
	\label{fig:DispersionGaps}
\end{figure}

A major challenge, however, is that the appearance of Chern bands appears to be very sensitive to model details. For example, Hartree-Fock band renormalization effects changes the Chern number of bands in trilayer graphene\cite{Zhang:2018wj}. The circumstances under which the hBN substrate gives rise to Chern bands in tBG are still not fully understood~\cite{Bultinck:2019wp,Zhang:2019hm}. With this in mind, in this article we carry out a detailed study of a novel twisted candidate material~\cite{LiuPRX2019,2019arXiv190500622M}: monolayer graphene twisted with respect to AB-stacked bilayer graphene, called twisted monolayer - bilayer graphene (tMBG).

We find that for realistic hopping parameters, and for a wide range of twist angles, this system hosts isolated flat bands separated by energy gaps from the remaining bands. In each valley, the two flat bands have a net nonzero Chern number, in accord with predictions of Ref.~\cite{LiuPRX2019} for twisted $N+M$ layers of graphene. The flat bands can be efficiently controlled by gates: a perpendicular electric field allows to tune both the dispersion of the flat bands and gaps between them. Even more interestingly, realistic electric fields can induce a series of topological transitions, accompanied by changes of Chern numbers of the flat bands, and redistribution of the Berry's curvature in the Brillouin zone. The topological character of the flat bands gives rise to a characteristic response to a perpendicular magnetic field, having clear signatures in the Hofstadter butterfly, which we compute. 

Finally, we show using Hartree-Fock calculations that interactions will favor spin- and/or valley-polarized states in a broad regime of twist angles. In particular, states with a filling factor $\nu=1$ are found to be both spin- and valley-polarized, exhibiting a QAH effect. Their stability in a broad range of angles, as well as tunability by gates, make tMBG a promising candidate for realizing robust QAH effect in a sample-independent manner. We note that previous works~\cite{Khalaf:2019ud, BernevigTrilayerMetal,MacDonaldTrilayer2019} investigated electronic properties of trilayer graphene where both top and bottom layer are rotated with respect to the middle one; such a system can host both flat bands~\cite{Khalaf:2019ud,MacDonaldTrilayer2019}, and a perfect metal~\cite{BernevigTrilayerMetal}.

{\em Continuum model ---} We first introduce a continuum model used to calculate the band structure for a monolayer graphene on top of Bernal-stacked bilayer graphene~\cite{McCann:2012ft,Jung:2014hj}. An untwisted graphene layer has lattice vectors ${\bf a}_1 = a (1,0)$ and ${\bf a}_2 = a(1/2, \sqrt{3}/2)$ with $a = 0.246$ nm. 
The reciprocal lattice vectors are ${\bf G}_1 = \frac{2\pi}{a} (1 , -1/\sqrt{3})$ and ${\bf G}_2 = \frac{2\pi}{a} (0 , 2/\sqrt{3})$.
The rotated layers have reciprocal lattice vectors ${\bf G}_{\ell,i} = R(\pm \theta/2)  {\bf G}_i$, where $R(\pm \theta/2)$ denotes rotation by $\pm \theta/2$, and we choose a positive twist angle for layer $\ell = 1$ and a negative one for layers $\ell = 2,3$. The reciprocal Moir\'{e} vectors are given by: 
\begin{equation}
	{\bf G}^M_i = R(- \theta/2) {\bf G}_i - R(+ \theta/2)  {\bf G}_i.
\end{equation}
The Moir\'{e} unit cell lattice vectors are defined by  the relation ${\bf L}^M_i \cdot {\bf G}^M_j = 2 \pi \delta_{ij}$, where the Moir\'{e} lattice constant is $|{\bf L}^M_i| = \frac{a}{2 \sin \theta/2}$. The $K$ and $K'$ points of the valley $\xi=\pm1$ are given by ${\bf K}^{(\xi)}_\ell = - \xi (2 {\bf G}_{\ell,1} + {\bf G}_{\ell,2})/3$. 

To describe the low-energy physics of the system, we employ the continuum model that extends the one developed for tBG graphene 
(see Ref.~\cite{Bistritzer:2011ho,PhysRevX.8.031087,2019arXiv190901545B}). We consider two graphene sheets (layers 1 and 2)  that comprise a tBG and add a third, AB-stacked graphene layer aligned with layer 2~\cite{SuarezMorell:2013et,LiuPRX2019,2019arXiv190500622M}. Retaining the couplings between adjacent layers, we arrive at  the following $6\times6$ Bloch Hamiltonian matrix for a single valley $\xi$ [in the basis of wave function amplitudes on A, B sublattices in the three layers $(A_1, B_1, A_2, B_2, A_3, B_3)$] 
\begin{equation}
	H^{(\xi)} = \begin{pmatrix}
		H_1 & U^\dagger & 0  \\
		U & H_2 & B^\dagger \\
		0 & B & H_3 
	\end{pmatrix},
\end{equation}
where in the continuum approximation the intralayer blocks are 
\begin{equation}
	H^{(\xi)} = \begin{pmatrix}
		0 & -v \pi^\dagger  \\
		-v \pi & 0 
	\end{pmatrix}
\end{equation}
where the $\pi$-terms, given by
\begin{equation}
    \pi = \left[ R(\pm \theta/2) ({\bf k} - {\bf K_\ell^{(\xi)}} ) \right]
		\cdot (\xi , i),
\end{equation}
depend on the twist angle of the layer, as well as the valley index $\xi$. The matrix $U$ is the effective local interlayer coupling between the twisted layers 1 and 2,
\begin{eqnarray}
	U ^{(\xi)}&=& \begin{pmatrix}
		u & u' \\ u' & u 
	\end{pmatrix} 
	+ \begin{pmatrix}
		u & u' \omega^{- \xi} \\ u' \omega^\xi & u 
	\end{pmatrix}e^{i \xi {\bf G}^M_1 \cdot {\bf r}} \nonumber \\ &&
	+ \begin{pmatrix}
		u & u' \omega^\xi \\ u' \omega^{-\xi} & u 
	\end{pmatrix} e^{i \xi ({\bf G}^M_1 + {\bf G}^M_2) \cdot {\bf r}}.
\end{eqnarray}
The matrix $B$ describes the coupling between layers  2  and 3, and, due to their crystalline alignment, is local in momentum: 
\begin{equation}
	B^{(\xi)} = \begin{pmatrix} 
		-v_4 \pi  & t_1 \\ -v_3 \pi^\dagger & -v_4 \pi, 
	\end{pmatrix}.
\end{equation}
Based on Refs.~\cite{Bistritzer:2011ho,PhysRevX.8.031087,Chen:2020tm} for the twisted bilayer parameters and \cite{Jung:2014hj} for the AB-stacked bilayer parameters, we set the hoppings $t = -2.61$ eV, $t_1 = 0.361$ eV, $t_3 = 0.283$ meV and $t_4 = 0.140$ eV. The Dirac velocities are related to the bare hoppings by $v_i = t_i \sqrt{3}a/2$. The Moir\'{e} hopping is $u' = 110$ meV and $u = 0.5 u'$. To model the effect of a hBN substrate, we include a $10$ meV sublattice symmetry breaking mass term on the first layer.\cite{Bultinck:2019wp}

{\em Flat bands and their topology ---} The lower symmetry of tMBG compared to the tBG implies that the Dirac cones in the flat bands are no longer symmetry-protected. Both at the ${\bf K}$-point of the mini-Brillouin zone (mBZ), corresponding to the position of the Dirac cone in the monolayer, as well as at the ${\bf K'}$-point, corresponding to the position of the band-touching in the AB stacked bilayer, we see a gap opening between the flat bands in a broad range of twist angles. 

This situation is markedly different from TBG, where (within the continuum approximation) the interlayer coupling keeps the Dirac cones intact and only renormalizes the Dirac velocity \cite{Bistritzer:2011ho,Song:2018ul}. Consequently, tMBG does not have a discrete sequence of 'magic angle' where the Dirac velocity vanishes. Nonetheless, we find that tMBG exhibits two flat bands in each valley, separated by a sizeable gap from other conduction and valence bands. The flat bands exists in a relatively wide range of angles from $\theta =0.88^\circ$ to $\theta = 1.5^\circ$, where their bandwidth  never exceeds 45 meV. Furthermore, in the range $1^\circ < \theta < 1.4^\circ$ their bandwidth is smaller than the gaps to other bands, as shown in Fig.~\ref{fig:DispersionGaps}a.

The band structure in this regime is illustrated Fig.~\ref{fig:DispersionGaps}d. Each valley contains two bands that overlap in a range of energy values. Interestingly, the lower band is topologically trivial whereas the top band has a nonzero Chern number $C = \pm 1$, with the sign being opposite for the two valleys.  The Chern numbers are computed using the method of Ref.~\cite{Fukui:2005cn}.  We note that the dispersion of the flat bands is significantly affected by the inclusion of smaller hopping parameters, however, their total Chern number conforms to the general expectations for twisted $N+M$ layer graphene~\cite{LiuPRX2019}. 

\begin{figure}[t]
    \includegraphics[width=\columnwidth]{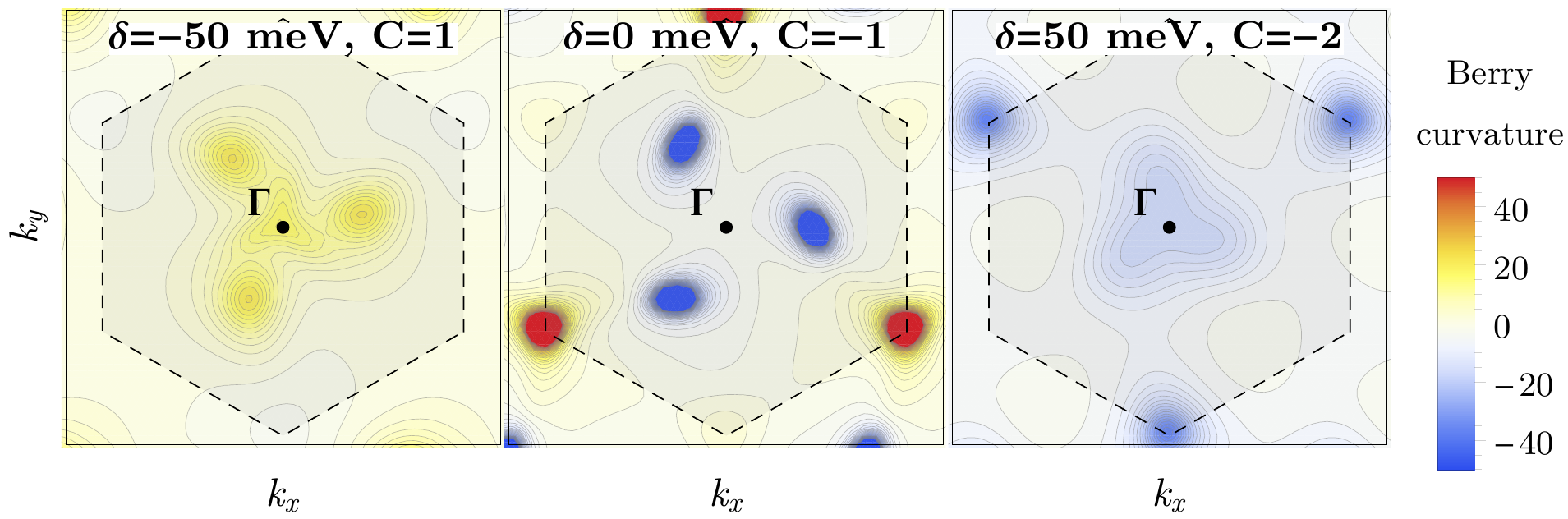}
	\caption{The Berry curvature in the mini-Brillouin zone for the upper flat band in the $\xi =1$ valley at a twist angle $\theta=1.08^\circ$. For $\delta = 0$ meV there are clear signatures of three gapped Dirac cones away from high-symmetry points, making the Berry curvature highly inhomogeneous throughout the mBZ. In contrast, at large fields $\delta$ the Berry curvature is more uniform and thus more analogous to a Landau level.}
	\label{Fig:Berry}
\end{figure}


The distribution of the Berry curvature for the flat bands in the mini-Brillouin zone is shown in Fig.~\ref{Fig:Berry}. Often, as is the case e.g. in the Haldane model~\cite{Haldane:1988gh}, the Berry curvature is concentrated in the vicinity of gapped Dirac points. Examination of the Berry curvature distribution for both trivial and Chern bands in tMBG indicates the presence of three gapped Dirac points at non highly symmetric points in the Brillouin zone. As a result, the Berry curvature is highly inhomogeneous in the BZ. 

{\em Tunability of flat bands by gates---} Next, we investigate the tunability of the flat bands by gates, finding that both their dispersion and topological properties can be efficiently controlled. The gate-induced displacement field is modeled by including a potential $-\delta/2$ on the first layer and $+\delta/2$ on the third layer. The resulting band structure for a particular value of twist angle, and bias $\delta=\pm 50\, {\rm meV}$, illustrated in Fig.~\ref{fig:DispersionGaps}c. and e., exhibits an indirect band gap between flat bands. Interestingly, this also leads to a more uniform distribution of the Berry's curvature in the mBZ (Fig.~\ref{Fig:Berry}).

The gap between conduction and valence bands (both direct and indirect) changes substantially as a function of applied perpendicular displacement field and twist angle $\theta$, see Fig.~\ref{fig:DispersionGaps}b. While changing the twist angle does not close the gap, under a change of displacement field the system undergoes several topological transitions. In the regimes where the conduction and valence band also have a positive indirect band gap, the conduction band has either $C=1$ for $D<0$, or $C = -2$ for $D>0$. This is consistent with the calculations of Ref.~\cite{Chen:2020tm}.

\begin{figure}[t]
	\includegraphics[width=\columnwidth]{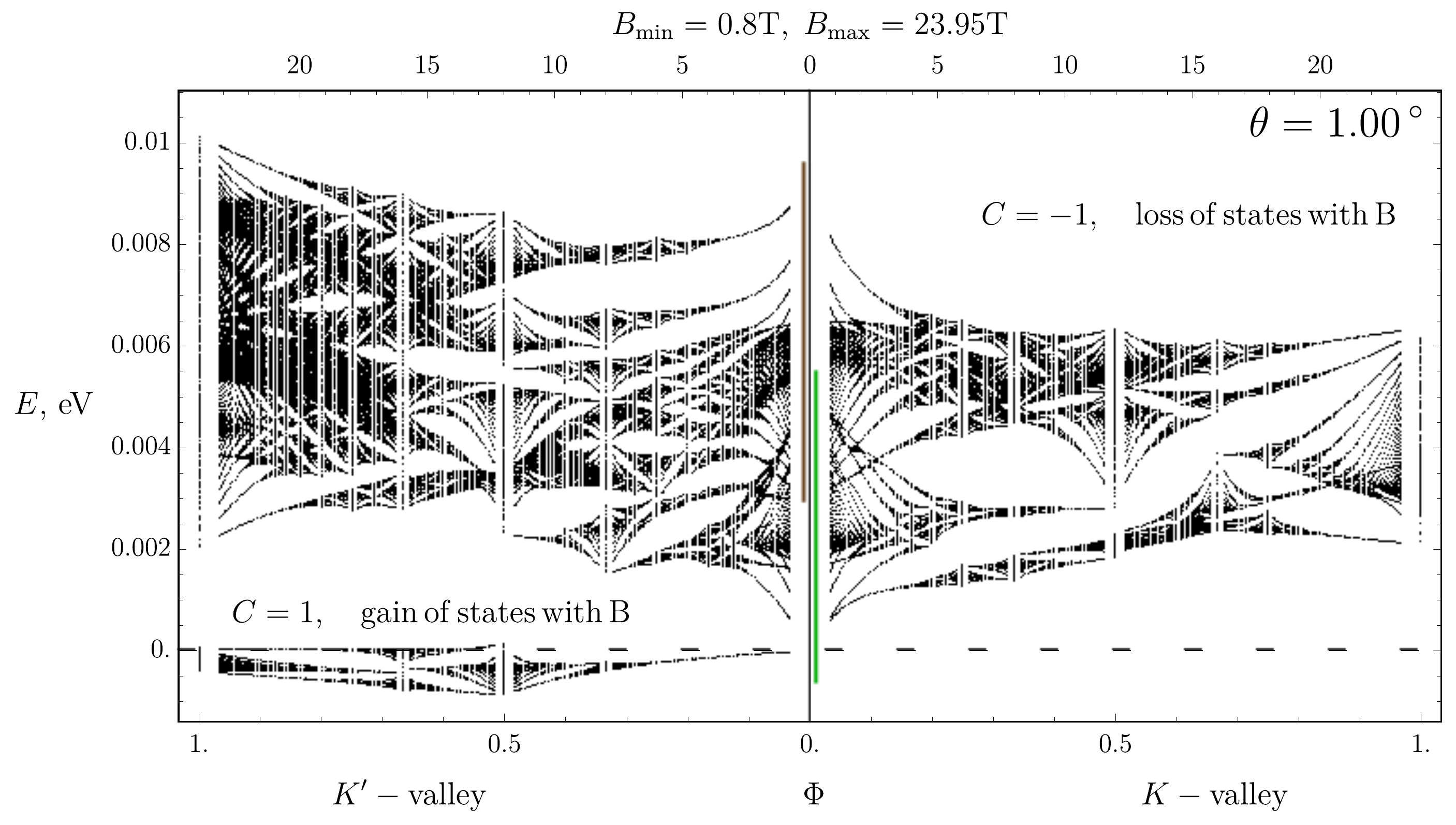}
	\caption{The energy levels of tMBG in a magnetic field, for valley $\xi =1$ and only the flat bands, without electric field ($\delta =0$ meV). An applied positive magnetic field clearly decreases the number of available states, as is expected due to the net Chern number $C=-1$. The response in the other valley $\xi=-1$ can be obtained by changing the sign of the applied field.}
	\label{Fig:Hofstadter}
\end{figure}

{\em Response to a magnetic field --- } A non-zero Chern number per valley gives rise to a distinct response to a perpendicular magnetic field. According to the Streda formula \cite{Streda:2000dw}, $C = \frac{\partial n}{\partial B}$ and therefore the magnetic field reduces (increases) the number of states in a band with a negative (positive) Chern number. For tMBG, this implies that a magnetic field causes valley polarization.

In order to quantify this effect, we computed the spectrum of tMBG in the presence of a magnetic field. Previous works~\cite{Bistritzer:2011fh,Lian:2018vs,Hejazi:2019kv,SenthilHofstadter} have developed complementary approaches to study the behavior of tBG in a magnetic field. We used the method introduced by Ref.~\cite{Hejazi:2019kv}, which amounts to projecting the interlayer coupling onto a basis set of single-layer Landau level wave functions~\footnote{See Supplemental Material for  a comprehensive account of the formalism and details of the numerical procedure.} 

In Fig.~\ref{Fig:Hofstadter} we display the energy spectrum of the system as a function of the  magnetic field (the ``Hofstadter butterfly''~\cite{Hofstadter:1976js}). The right and left halves of the plot describe the evolution of the spectrum in $K$- and $K^\prime$-valleys, respectively, in the positive magnetic filed. Flipping the sign of the field interchanges the valleys. 

We observe that the states in the flat bands remain well separated from the other bands. It is also clear that a positive field decreases, and a negative field increases the total number of states in the valley $C=-1$.  This flow of states is compensated by the opposite process in the $C=1$ valley.

\begin{figure}[t]
	\includegraphics[width=\columnwidth]{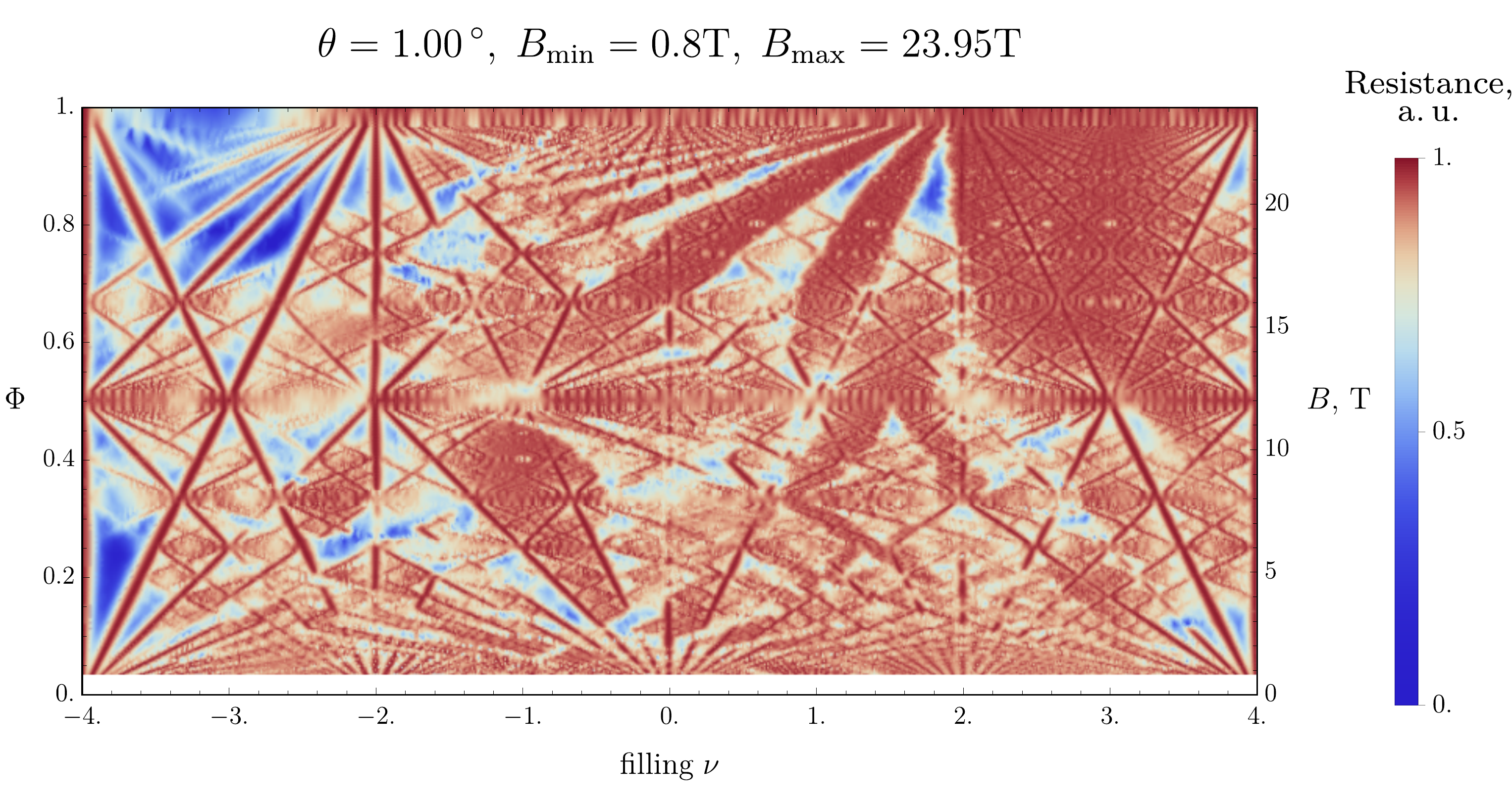}
	\caption{The density of states as a function of band filling $\nu$ and magnetic field $B$ in tMBG, also known as a Wannier plot. We show results up to one flux per Moir\'{e} unit cell, with $\delta = 0$ meV. The filling is measured relative to charge neutrality, so $|\nu|=4$ means complete filling or emptying of the flat bands. }
	\label{Fig:Wannier}
\end{figure}

We gain additional insight into the field response by converting the Hofstadter butterfly into a Wannier plot~\cite{1979PhRvB..19.6068C}, which shows the density of states
\footnote{In this work, limiting ourself to qualitative analysis,  we compute the density of states by phenomenologically  ascribing to the states in the  Landau levels   a finite lifetime (energy and flux independent). This procedure allows to highlight the insulating states on the $B-\nu$ phase diagram. Microscopic analysis of the elastic scattering processes in realistic samples is needed to make our predictions quantitative.} as a function of band filling and applied magnetic field, see Fig.~\ref{Fig:Wannier}. The full range of filling for the flat bands ranges from $\nu = -4$ to $\nu = 4$, and spin splitting is neglected. 

Recall that in a Wannier plot, trivial bands will generate simple Landau 'fans' of insulating states around the band edges. The presence of a Dirac cone gives rise to a special sequence of Landau fans with slope $\pm 1, \pm 3, \pm 5, \ldots$. Bands with nontrivial Chern numbers, on the other hand, lead to an asymmetric behavior due to the valley-field coupling, that becomes more pronounced as the field increases. This is clearly seen for tMBG, where the density of states is significantly higher close to $\nu = -4$, washing out the Landau fans originating from other integer fillings. We want to emphasize also the special nature of the inverted Landau fan around $\nu = 2$ and fields close to one flux quantum per Moir\'{e} unit cell, where rather than a fan of insulating states a fan of conducting states is observed. 

{\em Interaction effects ---} The tunable flat bands of tMBG provide a promising playground for interaction-induced states. To identify the range of angles most favorable to observe correlated insulators, we have performed a comparison of the Hartree-Fock energies of variational ground states at several integer filling factors. The screened Coulomb interaction $V(r) = \frac{e^2}{ 4 \pi \epsilon_0 \epsilon r} e^{-r / d}$, with screening length $d = 40$ nm and dielectric constant $\epsilon = 20$, is projected onto the upper flat bands' wave functions obtained from the continuum model discussed above. 

We focus on biased samples, with $\delta=\pm 50\, {\rm meV}$, since in this regime a gap between flat bands develops already at the single-particle level. Moreover, the distribution of Berry's curvature is relatively uniform, suggesting an analogy with Landau levels and the phenomenon of quantum Hall ferromagnetism that favors interaction-induced, spin- and/or valley-polarized states. 

In agreement with this intuition, our variational calculations indicate that at $\nu=1$, a fully spin- and valley-polarized state is energetically favored. Such a state is a quantum anomalous Hall (QAH) insulator. For $\delta=50\, {\rm meV}$, a QAH state at $\nu=1$ is found to be a ground state for a broad range of twist angles, $\theta\in (1^\circ;1.5^\circ)$, while for $\delta=-50\, {\rm meV}$ it is favored in a slightly smaller range, $\theta\in (1^\circ; 1.4^\circ)$. At $\nu=2$, we find two possible ground states: either fully valley polarized or fully spin-polarized. The fully spin-polarized state would not break time-reversal symmetry and thus would not show a QAH, but the spin-unpolarized state will show a QAH effect. At the variational level, these states are degenerate, and consideration of smaller, lattice terms is required to lift this degeneracy. 

{\em Outlook ---} To summarize, we studied the flat bands in a single layer graphene placed on top of an AB-stacked bilayer with twist. Including smaller interlayer hoppings modifies the dispersion of the bands while maintaining their total non-zero Chern number, first found within an idealized model of $N+M$ twisted graphene~\cite{LiuPRX2019}. 
The topological character of the bands persists for a broad range of twist angles. This, combined with a remarkable tunability of bands by an applied perpendicular field, makes TGBG a versatile platform for realizing controllable, topological bands, and experimentally studying the interplay of interactions and Berry curvature of electronic wave functions. 

Using variational Hartree-Fock calculations, we found that interactions favor spin- and/or valley-polarized states at integer filling factors. In particular, the spin and valley-polarized state at $\nu=1$ exhibits quantum anomalous Hall effect with a Hall conductivity $\sigma_{xy}=\pm e^2/h, \pm 2e^2/h$ that can be tuned by perpendicular electric field.  These physics of interaction-induced states in bands with non-zero Chern number carries certain similarity~\cite{SondhiChern} to quantum Hall ferromagnets, and in the future it would be interesting to study the effects of inhomogeneous distribution of Berry's curvature, as well as the energetics of spin and valley textures, such as skyrmions.

{\it Acknowledgements. --} We thank Amir Yacoby for inspiring discussions. This work was supported by the Swiss National Science Foundation via an Ambizione grant (L.~R.) and a regular project (D.~A. and I.~P.).

{\it Note added.--} While this manuscript was being prepared for submission, we have learned of two recent experiments~\cite{Chen:2020tm,Shi:2020uy} that report the observation of flat bands and signatures of interaction-induced insulating states in tMBG.

\end{document}